\documentclass[prx,twocolumn,showpacs,preprintnumbers,amsmath,amssymb,longbibliography]{revtex4-1}
\usepackage{graphicx}
\usepackage{amsmath}
\usepackage{url}
\usepackage{color}
\usepackage{placeins}
\usepackage{float}
\usepackage{chngcntr}
\restylefloat{table}
\DeclareGraphicsExtensions{.pdf,.eps,.png,.jpg,.mps}

\begin{document}

%\preprint{}

\title{Symmetry-protected metallic and topological phases in penta-materials}

\author{Sergio Bravo}
\affiliation{Departamento de F\'{i}sica, Universidad T\'{e}cnica
Federico Santa Mar\'{i}a, Casilla 110-V, Valpara\'{i}so, Chile}
\affiliation{Materials Science Factory, Instituto de Ciencia de Materiales de Madrid, Consejo Superior de Investigaciones Cient\'\i ficas, C/ Sor Juana In\'es de la Cruz 3, 28049 Madrid, Spain}

\author{J. D. Correa}
\affiliation{Facultad de Ciencias  B\'asicas, Universidad de Medell\' \i n, 
Medell\' \i n, Colombia}

\author{Leonor Chico}
\email{leonor.chico@icmm.csic.es}
\affiliation{Materials Science Factory, Instituto de Ciencia de Materiales de Madrid, Consejo Superior de Investigaciones Cient\'\i ficas, C/ Sor Juana In\'es de la Cruz 3, 28049 Madrid, Spain}

\author{M. Pacheco}
\email{monica.pacheco@usm.cl}
\affiliation{Departamento de F\'{i}sica, Universidad T\'{e}cnica
Federico Santa Mar\'{i}a, Casilla 110-V, Valpara\'{i}so, Chile}

\date{\today}

\begin{abstract}
We analyze the symmetry and topological features of a family of materials closely related to penta-graphene, derived from it by adsorption or substitution of different atoms. 
Our description is based on a novel approach, called topological quantum chemistry, that allows to characterize the topology of the electronic bands, 
based on the mapping between real and reciprocal space. 
 In particular, by adsorption of alkaline (Li or Na) atoms we obtain a nodal line metal at room temperature, with a continuum of Dirac points around the perimeter of the Brillouin zone. This behavior is also observed in some substitutional derivatives of penta-graphene, such as penta-PC$_2$. 
 Breaking of time-reversal symmetry can be achieved by the use of magnetic atoms; we study penta-MnC$_2$, which also presents spin-orbit coupling and reveals a topological insulator phase. 
 We find that for this family of materials, symmetry is the source of protection for metallic and nontrivial topological phases that can be associated to the presence of fractional band filling, spin-orbit coupling and time-reversal symmetry breaking.
\end{abstract}

%\keywords{}

\maketitle

\section{Introduction} 
 
Topological phases of materials due to spatial and non spatial
symmetries are the subject of enormous attention, both from the fundamental and the applied viewpoint. 
Firstly, this is because of the promising features related to the presence of robust
states at boundaries, such as protected surface states, and also due to the appearance of novel
quantum phenomena, showing unique signatures in the electronic transport, optical response
and other experimentally relevant magnitudes for applications \cite{Tse2016,Tse2010,Bauer2016,Wang2017}.

The search for materials with such desirable properties requires the concurrence of
symmetry reasoning along with {\it ab initio} calculations. On one side, symmetry
constrains in a clear and unambiguously way what kind of 
physical magnitudes are good quantum numbers to classify the states in the system, both
in direct and momentum space. On the other side, first-principles approaches allow for the 
quantitative characterization of the electronic structure of the material.
Combining these two basic ingredients, we can tailor properties of materials 
by design, in order to engineer topological nontrivial phases \cite{Bernevig2006,Hasan2010,Bansil2016}.

One of the most general and promising theoretical frameworks 
available for the study of novel materials 
is the so-called topological quantum chemistry (TQC) \cite{natBradlyn2017,Cano2018,Bradlyn2018}. This theory 
combines the non-local description of reciprocal space, in terms of bands, with the local, real space characterization employing atomic orbitals. 
It allows to classify the universal properties of all possible band structures of weakly correlated materials, making possible the identification of the topological
nature of their bands. 

In this work, we apply this novel approach to a family of two-dimensional (2D) materials related to penta-graphene (PG).
This theoretically predicted carbon allotrope
consists of a pentagonal, two-dimensional buckled lattice structure composed only of carbon atoms, first proposed by Tang {\it et al.} \cite{Tang2014} and later  by Zhang {\it et al.} \cite{Zhang2015}.

Lately, PG has received considerable attention from different
perspectives; it is a quasi-direct gap semiconductor that can be optimally combined with graphene and other 2D materials. Its potential applications have been lately explored \cite{Liu2016,Yuan2017,He2017,Rajbanshi2016,Chen2017,Krishnan2017,Bravo2018}, 
as well as the possibility of functionalization, adsorption and atomic substitution with
the aim of modifying its properties \cite{Wu2016,Li2016,QuijanoBriones2016,Enriquez2016,Xiao2016,Berdiyorov1,Berdiyorov2}. 

Some  of these modifications may change the semiconducting character of PG,
resulting in the appearance of metallic behavior. This feature has been presented in previous works  \cite{Enriquez2016,Berdiyorov2}; but most importantly, a general symmetry and topological study of these materials is still lacking. 
 
In what follows, we present a complete TQC analysis and {\it ab initio} calculations that explore the topological nature of these materials. 
We study their different phases upon inclusion of spin-orbit interaction and breaking of time-reversal symmetry. We find an evolution of the electronic band structure from a general nodal line located at the boundary of Brillouin zone (BZ), through a point-like Dirac node near Fermi level at the corner of the BZ, to topologically nontrivial phases in presence of spin-orbit coupling (SOC) plus time-reversal symmetry (TRS) breaking. These different phases are realized separately for specific penta-materials by means of first-principles calculations.

This work is organized as follows. We first perform an analysis for the space group of symmetry transformations present in this  family of materials. Next, a topological study is developed using the TQC approach. Certain characteristics of the materials, independent of their specific details, are derived from symmetry and topological analysis. Subsequently, we corroborate our study by means of density-functional theory calculations for several proposed penta-materials,  in order to illustrate the realization of the predicted phases. Finally, global conclusions pertaining this family of materials are drawn.
  
\begin{figure}[h!]
\centering
\includegraphics[width=0.55\columnwidth,clip]{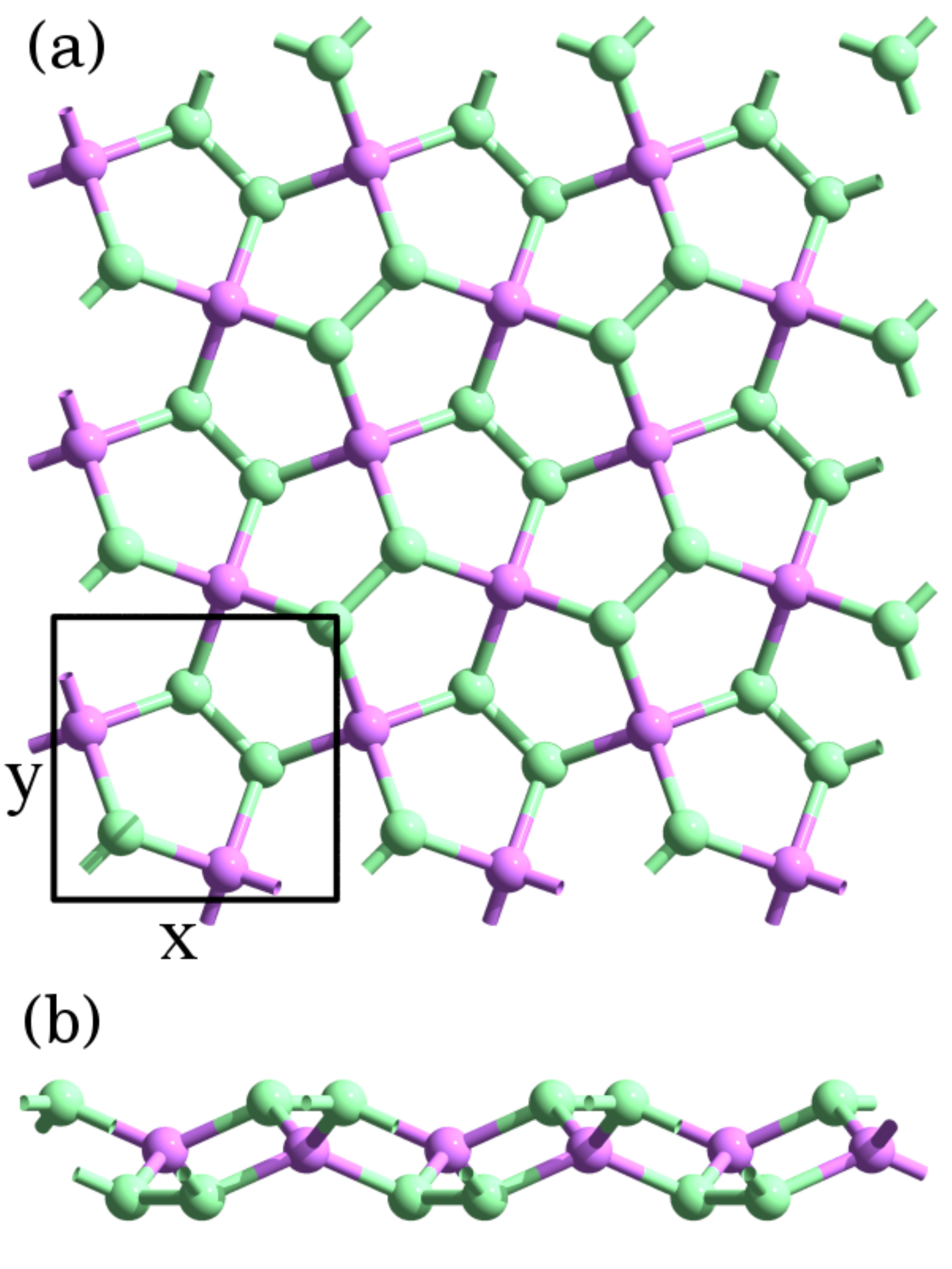}
 \caption{(a) Top and (b) side views of the PG lattice. The unit cell comprising 6 atoms is enclosed in a black square.}
  \label{figlattice}
\end{figure}

\section{Theoretical background}

We briefly summarize here the main concepts employed in the TQC 
approach in order to develop a topological classification of the electronic bands in penta-materials. 
First, a direct space characterization of the states is needed.  
This is done in terms of real space locations given by Wyckoff positions (WPs) and atomic or Wannier orbitals. Each WP has associated a site-symmetry group (SSG), composed by the set of symmetry operations that leaves the WP invariant. The SSG is a subgroup of the complete space group of the material \cite{Cano2018}. 
Assuming that there are $n_W$ orbitals at each WP, one has to identify the real-space irreducible representation with which these orbitals transform under the corresponding site-symmetry group. This gives a complete characterization of the states in real space. Next, the  connection to reciprocal space is achieved via a Frobenius procedure \cite{Cano2018}, which 
induces the so-called band representation \cite{Zak1982}, thus giving the relationship between real and momentum description.
If the WP under consideration is a maximal WP \cite{natBradlyn2017}, it has a maximal site symmetry group associated.
In this case the induced representation in reciprocal space is called an elementary band representation (EBR).
The EBRs are the building blocks to describe all the groups of bands for a particular space group \cite{Cano2018}. 
In order to identify the topological character of the bands, it is necessary to analyze their global connectivity in reciprocal space. By means of graph theory, all the possible connections among high-symmetry points compatible with the crystal symmetry constraints can be obtained \cite{Bradlyn2018,Vergniory2017}. This allows to classify the EBRs as connected or disconnected. Disconnected EBRs, which can be decomposed, present topological nontrivial behavior. The information about EBRs and compatibility relations for all space groups is available at the Bilbao Crystallographic server \cite{Bilbao}.

\section{Symmetry analysis}

Following the guidelines of the TQC approach, the fundamental starting point for the analysis of 
penta-materials is the knowledge of their symmetry group. 
The only formal requisite we impose to this class of materials is that their space group remains unaltered. 
Specifically, the space group of the penta-material lattice (see figure \ref{figlattice}) is given by 
the $P\bar{4}2_{1}mm$ or no. $113$ group \cite{MillerLoveBook}. 
Importantly, it is a nonsymmorphic group; it includes a glide plane with a fractional translation vector given by $\mathbf{t}=(a/2,a/2,0)$, where $a$ is the lattice constant of the material. The nonsymmorphic character has direct consequences on the energy bandstructure in momentum space, as we will see below. 

The topological nontrivial character of a material is directly related to the behavior of high-symmetry points (HSP) and high-symmetry lines (HSL). 
If a reciprocal wavevector $\bf k$ is a HSP or belongs to a HSL, there are certain operations of the space symmetry group $G$ (modulo a reciprocal wavevector) that leaves it invariant. This set of operations form the so-called little group of $\bf k$, $G_{\bf k}$, which is a subgroup of $G$ \cite{MillerLoveBook}. 
The two-dimensional BZ belonging to the space group no. 113 is presented in Fig. \ref{Wyckoff}.
The most relevant set of $k$ points in this group for the subsequent analysis are $\Gamma$, $X$, $M$, and the $Y$-line.

Let us begin with $\Gamma$.
All the transformations of the space group leave this point invariant, so its little group is isomorphic to the space group of the system. With respect to $X$, the symmetry transformations that leave the point invariant are $I$, $2_{001}$,$(2_{100}|\frac{1}{2}\frac{1}{2}0)$. Therefore, there are three equivalence classes for $G_X$, and in principle the same number of irreps. However, using the $\sum{h_{l}^{2}}=n$ constraint \cite{DresselJorioBook}, where $h_{l}$ is the dimension of the $l$-th irrep and $n=4$ is the order of the group, we obtain that only a two-dimensional irrep can exist. This irrep is labeled as $X_{1}$. 
For the $M$ point, the invariant operations are $I$, $2_{001}$, $\overline{4}^{+}_{001}$,$\overline{4}^{-}_{001}$,$(2_{100}|\frac{1}{2}\frac{1}{2}0)$,$(2_{010}|\frac{1}{2}\frac{1}{2}0)$,$(m_{110}|\frac{1}{2}\frac{1}{2}0)$ and $(m_{1\overline{1}0}|\frac{1}{2}\frac{1}{2}0)$. Thus $M$ is invariant under the complete space group $G$, so $G_M \cong G$.  
However, as $k\neq 0$, we have a  phase of $e^{i\mathbf{k_{M}\cdot t}}$ that is present in the wavefunction at momentum space. As  $\sum{h_{l}^{2}}=8$ it follows that $h_{1}=h_{2}=h_{3}=h_{4}=1$ and $h_{5}=2$. The character table for this little group is given in Table \ref{Mpointtable1}, that includes some complex characters due to the nonsymmorphic nature of the group. Finally, the symmetry transformations for the $Y$ line are $I$ and $(2_{010}|\frac{1}{2}\frac{1}{2}0)$. Since we have two classes with only one element each, there are two irreps. The corresponding character table is shown in Table \ref{Ypointtable1}, which also contains some complex characters.

\begin{table}[]
%\begin{ruledtabular}
%\renewcommand{\arraystretch}{1.3}
\begin{tabular}{@{\hspace{1em}}crrrcc@{\hspace{1em}}}
\toprule
M-point & $I$ & $2_{001}$ & $\overline{4}^{+}_{001}$ & $(2_{010}|\frac{1}{2}\frac{1}{2}0)$ & $(m_{110}|\frac{1}{2}\frac{1}{2}0)$ \\
\colrule
$M_{1}$ &  $1$ & $ 1$  & $1$ &\; \, $i$   & $-i$  \\
$M_{2}$ &  $1$ & $-1$  & $-1$ & $-i$  & $-i$ \\
$M_{3}$ &  $1$ & $-1$   & $1$ & \; \,$i$   & $-i$ \\
$M_{4}$ &  $1$ & $1$    & $-1$  & $-i$  &\;\,$i$\\
$M_{5}$ &  $2$ & $0$  & $0$ &\;  $0$   &\; $0$\\
\botrule
\end{tabular}
\caption{Character table for $G_M$.}
\label{Mpointtable1}
%\end{ruledtabular}
\end{table}

\begin{table}[]
%\begin{ruledtabular}
%\renewcommand{\arraystretch}{1.3}
\begin{tabular}{@{\hspace{1em}}c@{\hspace{1em}}r@{\hspace{1em}}c@{\hspace{1em}}}
\toprule
Y-line & $I$ & $(2_{001}|\frac{1}{2}\frac{1}{2}0)$  \\
\colrule
$Y_{1}$ &  $1$ &  $\omega$   \\
$Y_{2}$ &  $1$  & \, $\omega^{*}$   \\
\botrule
\end{tabular}
\caption{Character table for $G_Y$.  $\omega=e^{i\mathbf{k_{Y}\cdot t}}$}
\label{Ypointtable1}
%\end{ruledtabular}
\end{table}

The former data for the little groups allows for the description of the degeneracies at these HSPs and HSLs.
An important remark is pertinent in this place. 
Namely, if the system possesses time reversal symmetry, we have to resort to use only conjugate 
pairs of complex-valued irreps, known as physically irreducible representations 
\cite{natBradlyn2017}. With this in mind we analyze all points listed above. 
The $\Gamma$ point has one- and two-dimensional real representations, thus no TRS constraint is necessary.
At $X$ there is only one irrep with dimension two; therefore, this point always has a two-fold degeneracy for a spinless system. The $M$ point has four complex-valued representations and one real-valued representation.
TRS forces us to combine these four irreps in two pairs of conjugate physical irreps. This process yields the pairs $M_{1}+M_{4}$, $M_{2}+M_{3}$ and the real $M_{5}$. All three physical irreps are two-dimensional.
Thus, as long as TRS holds and the space group is nonsymmorphic, the energy bands at this point will be two-fold degenerate.  Finally,  imposing TRS at the $Y$ line
we are left with only one possible physical irrep, $Y_{1}+Y_{2}$, which is also two-dimensional. Therefore, every point located at $Y$ has a two-fold degeneracy.
It is worth to notice that the $Y$ line, along with the $X$ and $M$ points, comprise
all the inequivalent points at the BZ boundary. As there is a two-fold degeneracy in each case, 
it occurs a two-fold band touching over the entire BZ perimeter. This phenomenon is known as a nodal-line degeneracy \cite{Young2015}. 

If the spin degree of freedom is taken into account, the first trivial consequence is the doubling of the spinless original degeneracy. In this case, the trivial addition of spin yields a four-fold degeneracy along the nodal line. 
A more interesting scenario arises when SOC is included. Symmetry considerations must be extended to include double groups \cite{DresselJorioBook}. Consequently, the analysis for the little groups should be performed again, and the bands should be relabeled according to the new spinorial irreps.

Let us proceed with the analysis. The little group of the $\Gamma$ point, $G_\Gamma$, is enlarged to embrace two new spinorial irreps labeled by $\overline{\Gamma}_{6}$ and $\overline{\Gamma}_{7}$;  see Appendix \ref{chartab}  for the character tables of all the double groups used in this work, and Table \ref{tableB1} for this particular group. These two irreps are two-dimensional, which implies that the maximal degeneracy at this point is two.
As commented above, the $X$ point is described by a unique two-dimensional irrep without spin; the inclusion of SOC enlarges the character table. This can be easily seen using the basic relation $\sum{h_{l}^{2}}=6$, where the solution is given by 
$h_{1}=2$ and $h_{2}=h_{3}=h_{4}=h_{5}=1$. Thus, four new one-dimensional complex representations are added with respect to the spinless case. The character table for this group is presented in Table \ref{tableB2}. If TRS holds, these irreps are joined in conjugate pairs, giving two possible physical irreps, namely,  $\overline{X}_{2}+\overline{X}_{4}$ and $\overline{X}_{3}+\overline{X}_{5}$, both two-dimensional. This last result implies that degeneracy is lifted, as in the $\Gamma$ point, splitting the group of four bands into two pairs of bands. 
The double group for the $Y$ line has two more irreps, $\overline{Y}_{3}$ and $\overline{Y}_{4}$, presented in the character Table \ref{tableB3}. These two irreps are complex and one-dimensional. Like in the previous cases, TRS implies the pairing of both irreps in the physical irrep $\overline{Y}_{3}+\overline{Y}_{4}$, forming a two-dimensional irrep. Therefore, as in $\Gamma$ and $X$, the possible four-fold degeneracy is lifted, yielding two stick-together, two-fold degenerate bands, along the whole line. 
finally for the $M$ point, the double group includes now two new spinorial irreps $\overline{M}_{6}$ and $\overline{M}_{7}$, both two-dimensional. The character table for $G_M$ is presented in Tables \ref{tableB4} and \ref{tableB5}. Again, we have complex-valued irreps. Under TRS these two irreps have to be paired in a single physical irrep, denoted as $\overline{M}_{6}+\overline{M}_{7}$. This irrep is four-dimensional; being the only option for the spinful case, we conclude that the $M$ point is unaffected by the inclusion of SOC, maintaining the four-fold degeneracy for the energy bands.
In summary, we have shown, based only on symmetry grounds, that the boundary nodal line disappears under SOC, leaving only a point-like degeneracy at $M$. 

We finish the exploration of symmetries in penta-materials by relaxing time-reversal invariance. If TRS is broken, single complex irreps can be physical representations without the need of coupling them in conjugate pairs. This has straightforward implications in the degeneracy landscape of the energy bands, with or without SOC. 
If TRS is absent and no SOC in considered, the following consequences can be deduced: (i) The $\Gamma$ point is still four-fold degenerated; (ii) The $X$ point becomes non-degenerate;  (iii) The $M$ point changes its degeneracy from four-fold to two-fold; and finally, (iv) at the $Y$ line we find non-degenerate bands, implying the disappearance of the nodal line for this case.

Additionally, in the SOC plus TRS breaking case we can deduce the following: (i) at $\Gamma$ nothing happens, since all irreps are already real; however,  (ii)  the conjugate pairs formed at $X$ under TRS break apart in the single complex one-dimensional irreps $\overline{X}_{2}$, $\overline{X}_{4}$, $\overline{X}_{3}$, $\overline{X}_{5}$.  Therefore, all bands are non-degenerate at this point. The $M$ point, which had a protected four-fold degeneracy, due to time-reversal and nonsymmorphic symmetries, ends up with a pair of two-fold degenerated bands. Finally, the degeneracy of the $Y$ line is lifted, leaving four non-degenerate bands for each group of the eight bands occurring in the spinful model.       
   
\section{Topological analysis}
\label{sec:topan}

\begin{figure}[h!]
\centering
\includegraphics[width=1.0\columnwidth,clip]{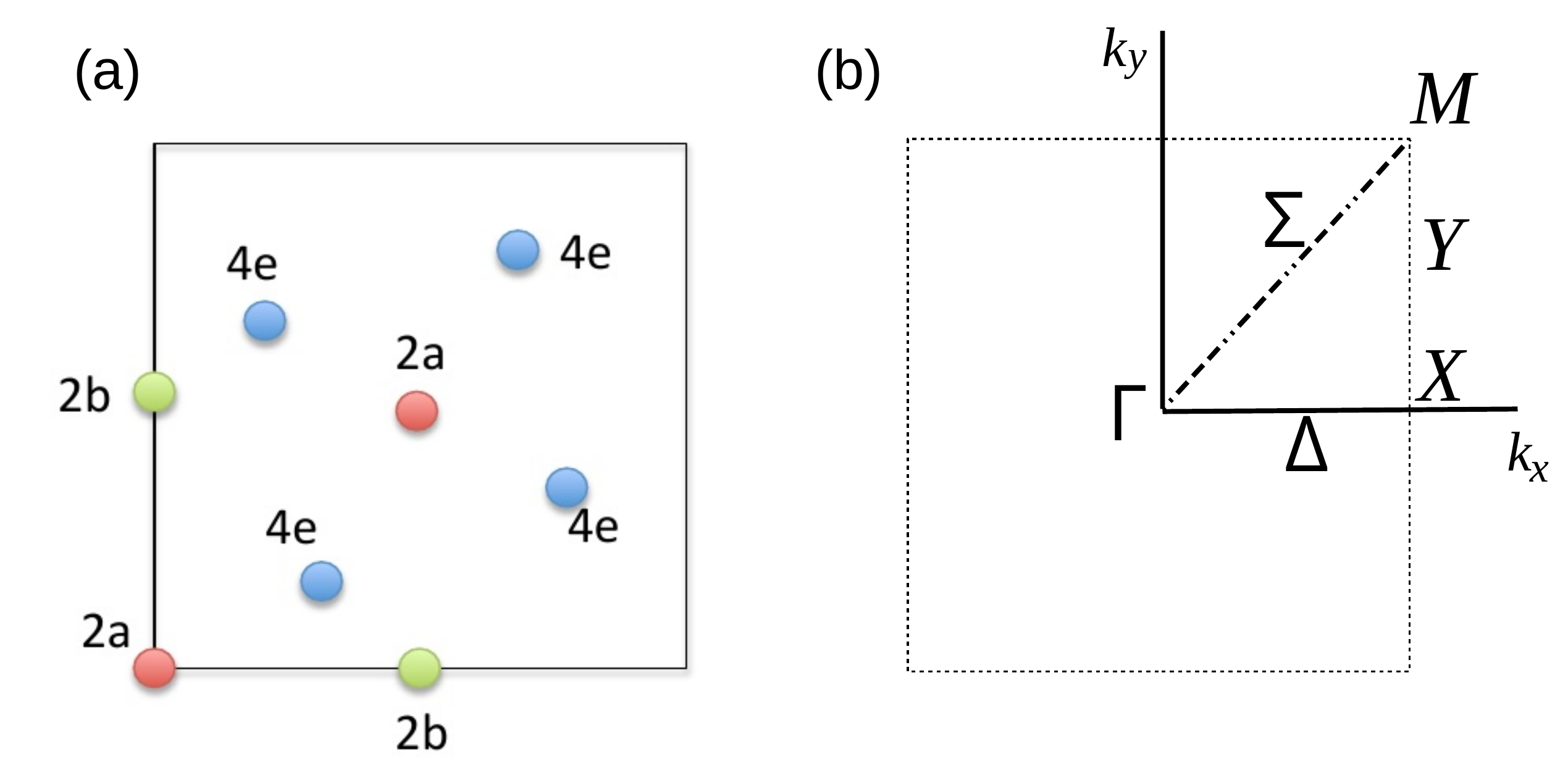}
 \caption{ a) Two-dimensional projection of the Wyckoff positions for the space group of penta-materials. b) Two-dimensional Brillouin zone for penta-materials.}
  \label{Wyckoff}
\end{figure}

In order to apply the topological analysis based on the symmetry description given in the previous section, we need to establish a model for the relevant energy range, namely, the vicinity of the Fermi level.
All the penta-materials presented here have the pentagonal lattice of PG as a basic 
structure (Fig. \ref{figlattice}), which has six atoms in its unit cell, four of them with coordination 3 and the other two with coordination 4.
In terms of Wyckoff positions (WPs), the atoms with coordination 3 are located at a
non maximal 4e WP, and those with coordination 4 are allocated at a maximal 2a WP. Figure \ref{Wyckoff} presents a graphical description of WPs for this particular space group.

\begin{table}[]
%\begin{ruledtabular}
\renewcommand{\arraystretch}{1.5}
\begin{tabular}{@{\hspace{1em}} c@{\hspace{1em}}r@{\hspace{1em}}r@{\hspace{1em}}r@{\hspace{1em}}r@{\hspace{1em}} }
\toprule
4e & $I$ & $2_{001}$ & $I^{d}$ & $2^{d}_{001}$ \\
\colrule
$A'$ &  $1$ & $ 1$  & $1$ & $1$  \\
$A'$ &  $1$ & $-1$  & $1$ & $-1$  \\
$\overline{E}_{1}$ &  $1$ & $-i$  & $-1$ & $i$  \\
$\overline{E}_{2}$ &  $1$ & $i$  & $-1$ & $-i$  \\
\botrule
\end{tabular}
\renewcommand{\arraystretch}{1}
\caption{Character table for the double site-symmetry group at 4e Wyckoff position.}
\label{latabla}
%\end{ruledtabular}
\end{table}

 With the WPs of the atoms identified, the real space description is completed enumerating the orbital components for each atomic site. The most important contribution for the considered penta-materials comes from the $p_{z}$ orbitals, and in particular from atoms at the 4e WP. We present a minimal model for the topological analysis based on the four $p_{z}$ orbitals at the 4e WP.   
Extensions to this model, including additional atoms, either adsorbed or as substitutions, can be also important and may involve other combinations of atomic orbitals. 
However, in terms of the essential topological behavior the main results are not modified, so we rely on this minimal model and discuss the necessary additions when appropriate.
 
Once the real space model is complete, it has to be translated to momentum space, with the aim to compute its induced band representation and the subsequent topological characteristics. To calculate the corresponding band representation, 
the site-symmetry group (SSG) related to the 4e WP \cite{Cano2018} should be identified. This group is composed of two operations, $I$, and $2_{001}$. 
Its character table is given in Table \ref{latabla} (we are only considering the first two columns and rows in this table for the single-valued SSG). 
In a first stage we ignore spin-orbit coupling and assume that TRS holds, which implies the use of physical irreducible representations \cite{natBradlyn2017}. The $p_{z}$ orbitals transform as the $A'$ irrep of this SSG. 
This information allows to define an induced band representation which gives as a result the symmetry (irrep labels) of the four bands throughout the entire BZ, as shown in Table \ref{table4}. Here we only show explicitly the TR-symmetric points $\Gamma, X$ and $M$. 

\begin{table}[]
%\begin{ruledtabular}
\renewcommand{\arraystretch}{1.5}
\begin{tabular}{@{\hspace{1em}} crrrrr}
\toprule
${\rm BZ \,\, point}$ \; \; \; & $A^{\prime}\uparrow G$ \; \; \\
\colrule
$\Gamma$ & $\Gamma_{1}(1)\oplus\Gamma_{3}(1)\oplus\Gamma_{5}(2)$ \; \; \\
$X$ & $2X_{1}(2)$  \; \; \\
$M$ & $M_{1}(1)M_{3}(1)\oplus M_{5}(2)$ \; \;\\
\botrule
\end{tabular}
\renewcommand{\arraystretch}{1}
\caption{Band representation for $4e$ WP with TRS and no SOC.}
\label{table4}
%\end{ruledtabular}
\end{table}

 A straightforward observation is that this band representation is composite \cite{natBradlyn2017}. 
This is to be expected, since our model is based on $p_{z}$ orbitals located at non-maximal WP. 
Nevertheless, we can express this band representation as a sum of EBRs 
coming from maximal WP: $A^{\prime}\uparrow G=(2a)\uparrow G\oplus(2c)\uparrow G$ (see \cite{Bilbao} for the complete list of EBRs for the group). 
The most important conclusion for this model is that all sets of bands are two-connected, and therefore all bands are topologically trivial. Particular examples of this phase show some variations of the electronic character of the material, depending on the specific band filling. The inclusion of SOC can lead to the appearance of additional phases in these materials. This implies the use of a double
group (double SSG) description, as mentioned before. The character table for the corresponding double group 
is given in Table \ref{latabla}. Due to TRS, we have to apply the conjugate pair procedure and join 
the $\overline{E}_{1}$ and $\overline{E}_{2}$ irreps in a single $\overline{E}_{1}+\overline{E}_{2}$ physical irrep for the spinful orbitals. This 
two-dimensional irrep induces a band representation in reciprocal space shown in Table \ref{table5}.

\begin{table}[h!]
%\begin{ruledtabular}
\renewcommand{\arraystretch}{1.5}
\begin{tabular}{@{\hspace{1em}} crrrrr }
\toprule
${\rm BZ \,\, point}$ \; \;  \; & $\overline{E}_{1}+\overline{E}_{2}\uparrow G$ \;\; \\
\colrule
$\Gamma$ & $2\overline{\Gamma}_{6}(2)\oplus2\overline{\Gamma}_{7}(2)$ \; \\
$X$ & $2\overline{X}_{2}\overline{X}_{5}(2)\oplus2\overline{X}_{3}\overline
{X}_{4}(2)$ \; \\
$M$ & $2\overline{M}_{6}\overline{M}_{7}(4)$ \; \\
\botrule
\end{tabular}
\renewcommand{\arraystretch}{1}
\caption{Band representation for $4e$ WP, with SOC and TRS.}
\label{table5}
%\end{ruledtabular}
\end{table}

 The above band representation takes into account eight bands arising from the spin degree of freedom.
Additionally, by exploring the character of all the EBRs with TRS for this double group, it can be verified 
that all sets of bands are connected, with a maximum of 4-connected bands (see \cite{Bilbao}). Thus, all bands are trivial in this case. Still, we have some SOC-induced transitions at the HSP and HSL that modify the degeneracy order as mentioned in the symmetry analysis, this has consequences on the electronic properties of particular penta-materials (see next Section).
 
Finally, we consider TRS breaking such that complex-valued irreps are allowed. With the same induction procedure employed above, we found the band representation shown in Table \ref{table6}.

\begin{table}[h!]
%\begin{ruledtabular}
\renewcommand{\arraystretch}{1.5}
\begin{tabular}[c]{@{\hspace{1em}} crrrrr }
\toprule
${\rm BZ \,\, point}$ \; \; & $2\overline{E}_{1}\uparrow G$ \; \; \\
\colrule
$\Gamma$ & $2\overline{\Gamma}_{6}(2)\oplus2\overline{\Gamma}_{7}(2)$ \; \\
$X$ & $2\overline{X}_{2}(1)\oplus2\overline{X}_{5}(1)\oplus2\overline{X}%
_{3}(1)\oplus2\overline{X}_{4}(1)$ \; \\
$M$ & $2\overline{M}_{6}(2)\oplus2\overline{M}_{7}(2)$ \; \\
\botrule
\end{tabular}
\renewcommand{\arraystretch}{1}
\caption{Band representation for $4e$ WP with SOC and no TR.}
\label{table6}
%\end{ruledtabular}
\end{table}

This is a composite band representation formed by two groups of four bands. We study only one group, since the other one has exactly the same structure. 

As it is well-known, degeneracy is lowered by TRS breaking and,  this is reflected in the band representation which becomes decomposable, a signal for the presence
of a topological set of bands \cite{natBradlyn2017,Cano2018,Bradlyn2018}. If an EBR is decomposable,
then different connectivity paths can appear among the  high symmetry points and lines through the BZ, which implies different topological phases in the material. The different topological realizations of the band representation correspond to  all  possible solutions of the compatibility relations between HSP and HSL over the BZ. We have carried this process for a two-dimensional BZ of the space group of penta-materials, finding the connectivity solutions presented in Table  \ref{table7} below.

\begin{table}[h!]
%\begin{ruledtabular}
\renewcommand{\arraystretch}{1.2}
\begin{tabular}[c]{@{\hspace{1em}} c  @{\hspace{3em}} c @{\hspace{1em}}}
%\begin{tabular}[c]{ c @{\qquad} @{\qquad} c }
\toprule
%\begin{align*}
%&
%[c]{cc}
HSP path   & Character
%\end{tabular}
\\
\hline
\hline
%&
%\begin{tabular}
%[c]{cc}
$\Gamma_{6}\rightarrow X_{2}\oplus X_{4}\rightarrow M_{6}$ & trivial\\
$\Gamma_{7}\rightarrow X_{3}\oplus X_{5}\rightarrow M_{7}$ & topological\\
\hline
%\end{tabular}
%\\
%&
%\begin{tabular}
%[c]{cc}
$\Gamma_{6}\rightarrow X_{3}\oplus X_{5}\rightarrow M_{6}$ & topological\\
$\Gamma_{7}\rightarrow X_{2}\oplus X_{4}\rightarrow M_{7}$ & trivial\\
\hline
%\end{tabular}
%&
%\begin{tabular}
%[c]{cc}
$\Gamma_{7}\rightarrow X_{2}\oplus X_{4}\rightarrow M_{6}$ & topological\\
$\Gamma_{6}\rightarrow X_{3}\oplus X_{5}\rightarrow M_{7}$ & trivial\\
%\end{tabular}
%\\
%&
%\begin{tabular}
%[c]{cc}
\hline
$\Gamma_{7}\rightarrow X_{3}\oplus X_{5}\rightarrow M_{6}$ & trivial\\
$\Gamma_{6}\rightarrow X_{2}\oplus X_{4}\rightarrow M_{7}$ & topological\\
%\end{align*}
\botrule
\end{tabular}
\renewcommand{\arraystretch}{1}
\caption{Topological phases.}
\label{table7}
%\end{ruledtabular}
\end{table}

\begin{figure}[h!]
\centering
\includegraphics[width=\columnwidth,clip]{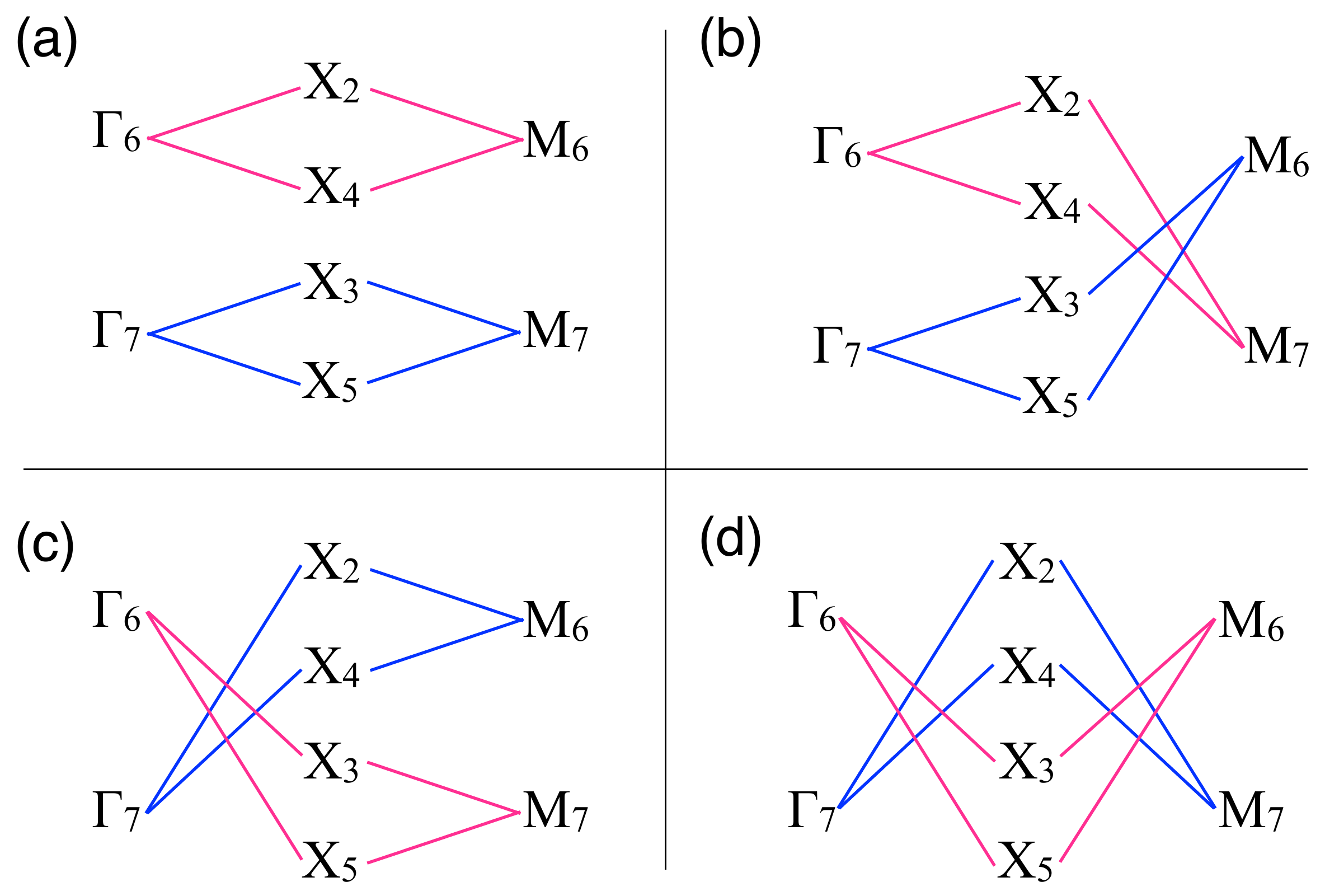}
 \caption{Scheme of the energy ordering  of bands (irreps) to some particular sequence at HSP and HSL.}
  \label{TPclip}
\end{figure}

These sets of bands have to be compared with the EBRs that the space group induces in momentum space.
As a general rule, if a band representation can be expressed as the combination of some EBRs, 
then the set of bands is trivial \cite{Cano2018}. Comparing the results for 
the band representations above with all possible ERBs, it can be seen that some sets cannot be expressed in terms of EBRs;  in conclusion, such bands are topological. The labeling presented in the  
Table \ref{table7} depicts this situation.

There are four different possible connectivities that depend on the particular characteristics and band filling of the material. 
We can build a toy model to grasp the general behavior of these phases by fixing the energy ordering 
of bands (irreps) to some particular sequence at HSP and HSL. The possible outcomes are depicted graphically in Fig. \ref{TPclip}.
There are three phases presenting nodal degeneracy and one phase with gapped character. 
The knowledge of the specific band filling is necessary in order to classify this topological behavior as metallic or insulating. This is strongly material-dependent, making it necessary to analyze specific cases. In particular, for penta-materials studied in this work, there is an interplay of metallic phases with or without nodes, along with electron or hole pockets near the Fermi energy, an scenario that has been found before \cite{Burkov2011}.  
In summary, if TRS is preserved and no SOC is included, penta-materials possess a general band structure with trivial bands in all its energy range, displaying a perimeter nodal line. The inclusion of SOC while maintaining TRS yields also trivial bands, but some degeneracies are lifted in the BZ, changing the character of the electronic properties. Breaking TRS with SOC produces a decomposable band representation that gives rise to four different topological phases,
according to the distinct possibilities for the band connectivity. 

This exhausts our study of the electronic band structure for penta-materials within a general group theory framework. Naturally, other perturbations could be included in order to modify the symmetry character of the underlying lattice with the possible induction of more topological phases.  

In the following section we apply this general group-theoretical description to some specific 
penta-materials. This is done with the aid of first-principles calculations and effective models. 

\begin{figure}[h!]
\centering
\includegraphics[width=1.0\columnwidth,clip]{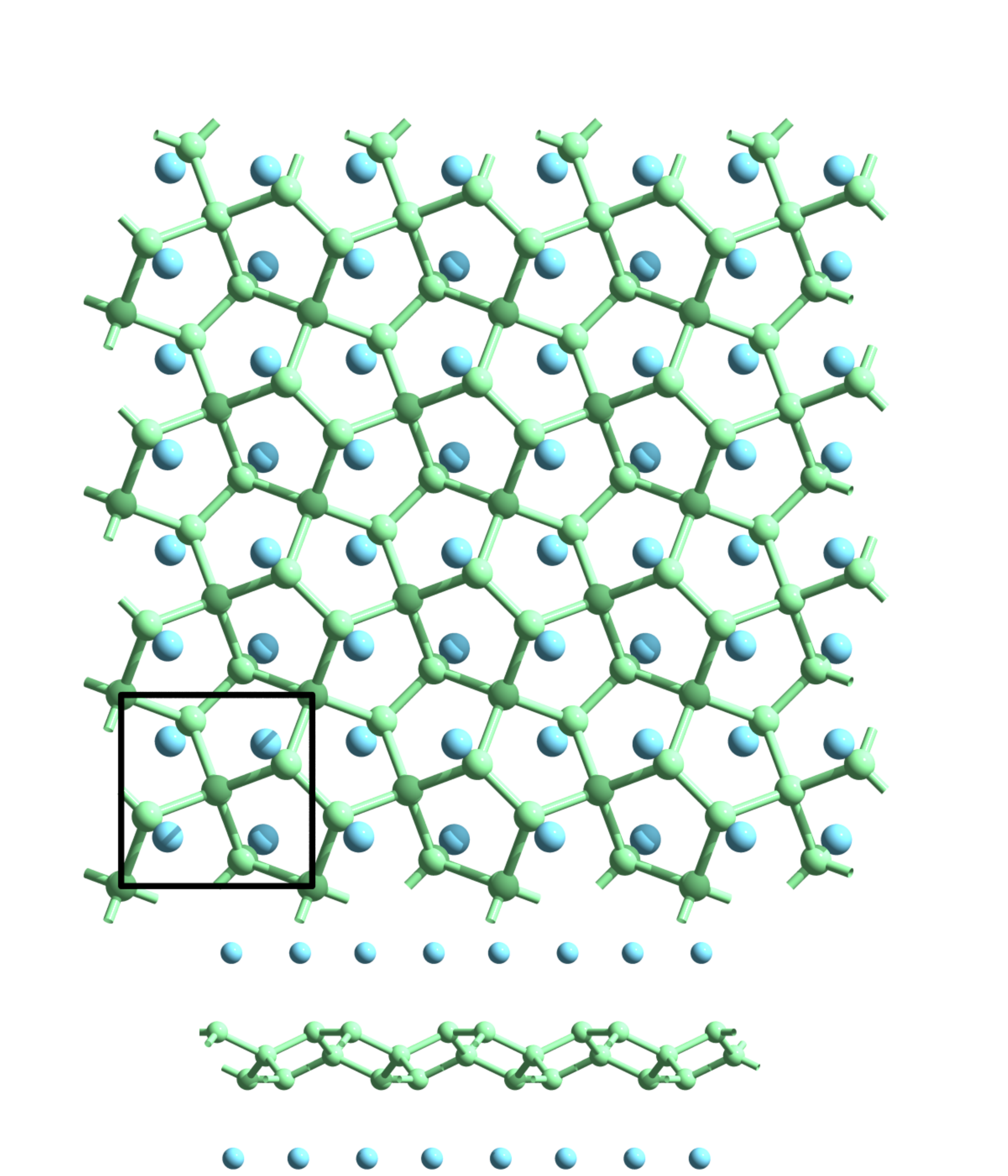}
 \caption{Li-PG relaxed lattice.}
  \label{LiPG}
\end{figure}

\section{Examples of penta-materials}

The first instance of this family of materials is penta-graphene. It has an all-carbon lattice structure and an insulating character. Two conduction bands and two valence bands are the main contributors to the low-energy range \cite{Zhang2015}. Applying the TQC analysis it can be concluded  that, since carbon has a weak SOC and TRS holds, PG is a trivial (band) insulator. In this context, PG presents little interest  due to its sizeable optical gap.
However, as mentioned before, we can explore additional configurations by functionalization, adsorption or atom substitution of penta-graphene without altering its original symmetry. This can be reinforced by an electron filling analysis as presented in \cite{Watanabe2016}. For the PG space group (No. 113) the band insulator filling is dictated by a $4n$ relation, where $n$ is a positive integer. For PG this yields a band filling of 36, which results in a band insulator state. If we substitute some carbon atoms we deviate from the $4n$-band filling, accomplishing a condition for the filling-enforced formation of a nodal (semi)metal. This has to be additionally cross-checked with a chemical stability study of the material \cite{Chen2018}.
  
We present first-principles calculations for several penta-materials based on PG, with an emphasis in both, global and local features of their energy bands. We explain the modifications performed in PG to achieve the specific penta-material and the particular phase realizations with respect to our previous symmetry and topological analysis.  

Our calculations were carried out in the density functional theory (DFT)  framework using SIESTA \cite{Soler2002}  and  Quantum ESPRESSO \cite{QE-2009,QE-2017} {\it ab initio} packages. The energy cutoff for the basis set was $80$~Ry~ for Quantum ESPRESSO and for SIESTA  calculations we employed localized atomic orbitals as a basis set (double-$\zeta$, single polarized).
  In  both  codes were employed norm-conserving  pseudopotentials and the structures were relaxed until the forces on the atoms were less than $0.04$~eV/\AA. Exchange-correlation was considered within the generalized gradient approximation (GGA), as proposed by Perdew, Burke, and Ernzerhof \cite{PBE1996}. The convergence of the total energy is ensured with a Monkhorst-Pack k-grid of $15\times 15\times 1$ in both cases. All the  geometrical parameters for the penta-materials presented below are summarized in Appendix \ref{latt} (see Table \ref{tableA1}).
In materials without spin-orbit coupling calculations were carried with SIESTA and Quantum ESPRESSO giving similar results. Calculations with spin-orbit coupling were performed exclusively with Quantum ESPRESSO.

\subsection{Symmetry-protected metallic phases}

\begin{figure}[h!]
\centering
\includegraphics[width=1.0\columnwidth,clip]{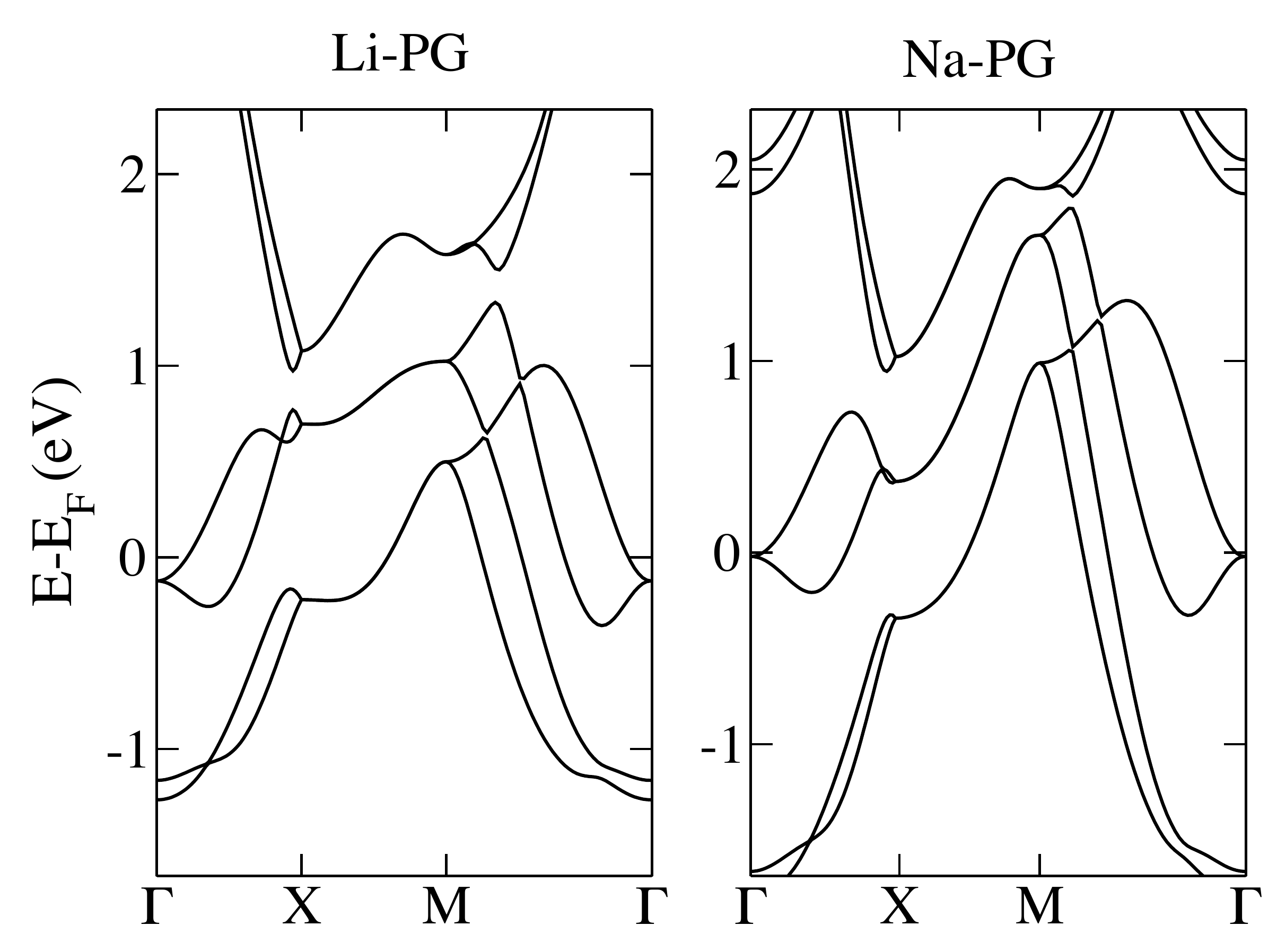}
 \caption{Band structures of Li and Na  absorbed on penta-graphene.}
  \label{pentamatbands1}
\end{figure}

\begin{figure}[h!]
\centering
\includegraphics[width=1.0\columnwidth,clip]{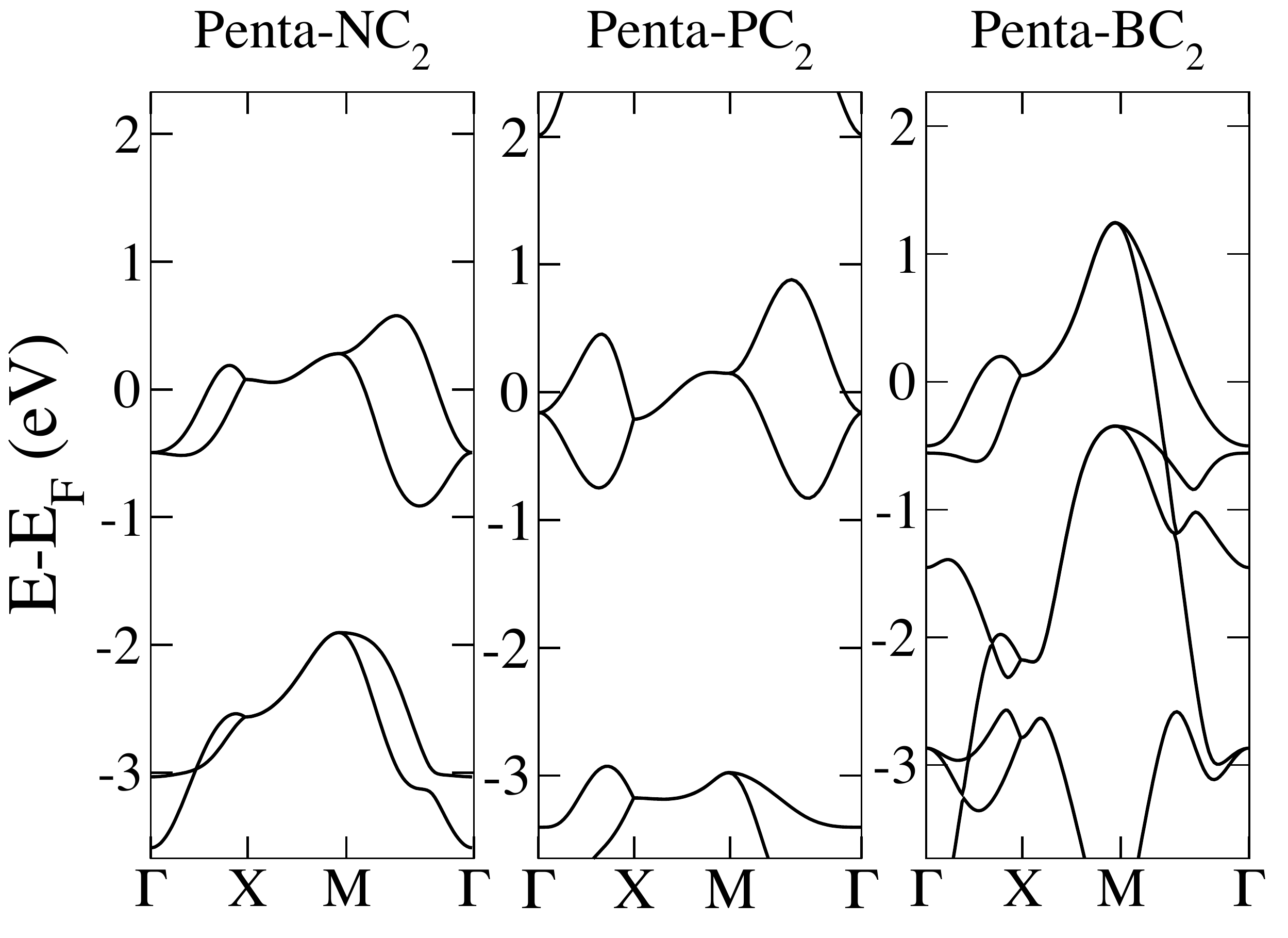}
 \caption{Band structures of  different XC$_2$   penta-materials.}
  \label{pentamatbands2}
\end{figure}

In order to access the metallic phases, i.e., to shift the conduction or valence bands, other elements rather than carbon should be added to PG. 
We first functionalize PG with adsorption of metallic atoms at 4e WP. This case has been previously explored for various elements, showing metallization of PG \cite{Xiao2016,Enriquez2016}. 
An example of a relaxed lattice structure with adsorbed Li is presented in Figure \ref{LiPG}. 
Also, electronic band structure calculations are shown for this case of Li-adsorbed PG (Li-PG) as well as Na-adsorbed PG (Na-PG) in Fig. \ref{pentamatbands1}.
Another possibility is to explore substitutional derivatives of PG, respecting the original symmetry.
 Particularly interesting for this work is the 2a WP, which corresponds to coordination-4 atoms, forming a penta-XC$_{2}$ configuration \cite{QuijanoBriones2016,Berdiyorov1,Berdiyorov2}, 
where X=\{B, N, P, Si, G\}. The lattice structure of these materials is exactly the same as PG, with modifications in the relative bond magnitudes and lattice constant. We show the band structures for X={B, N, P} in Fig. \ref{pentamatbands2}.

\begin{figure}[h!]
\centering
\includegraphics[width=1.0\columnwidth,clip]{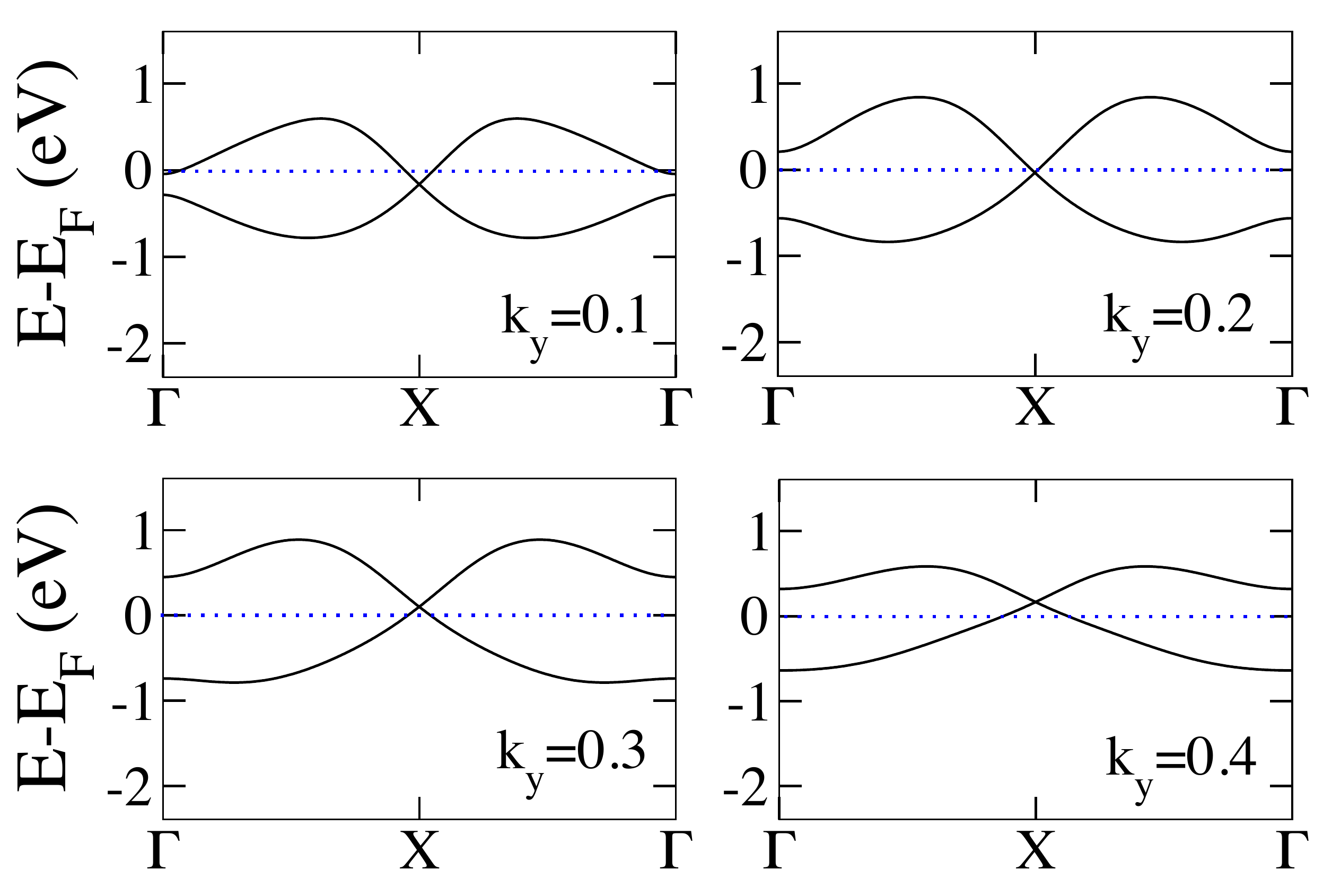}
 \caption{Y-line vicinity band structure of penta-NC$_{2}$ along different $k_{y}$ values, where $k_{y}\in(0,0.5)$.}
  \label{nodal}
\end{figure}

\begin{figure}
%[h!]
\centering
\includegraphics[width=1.0\columnwidth,clip]{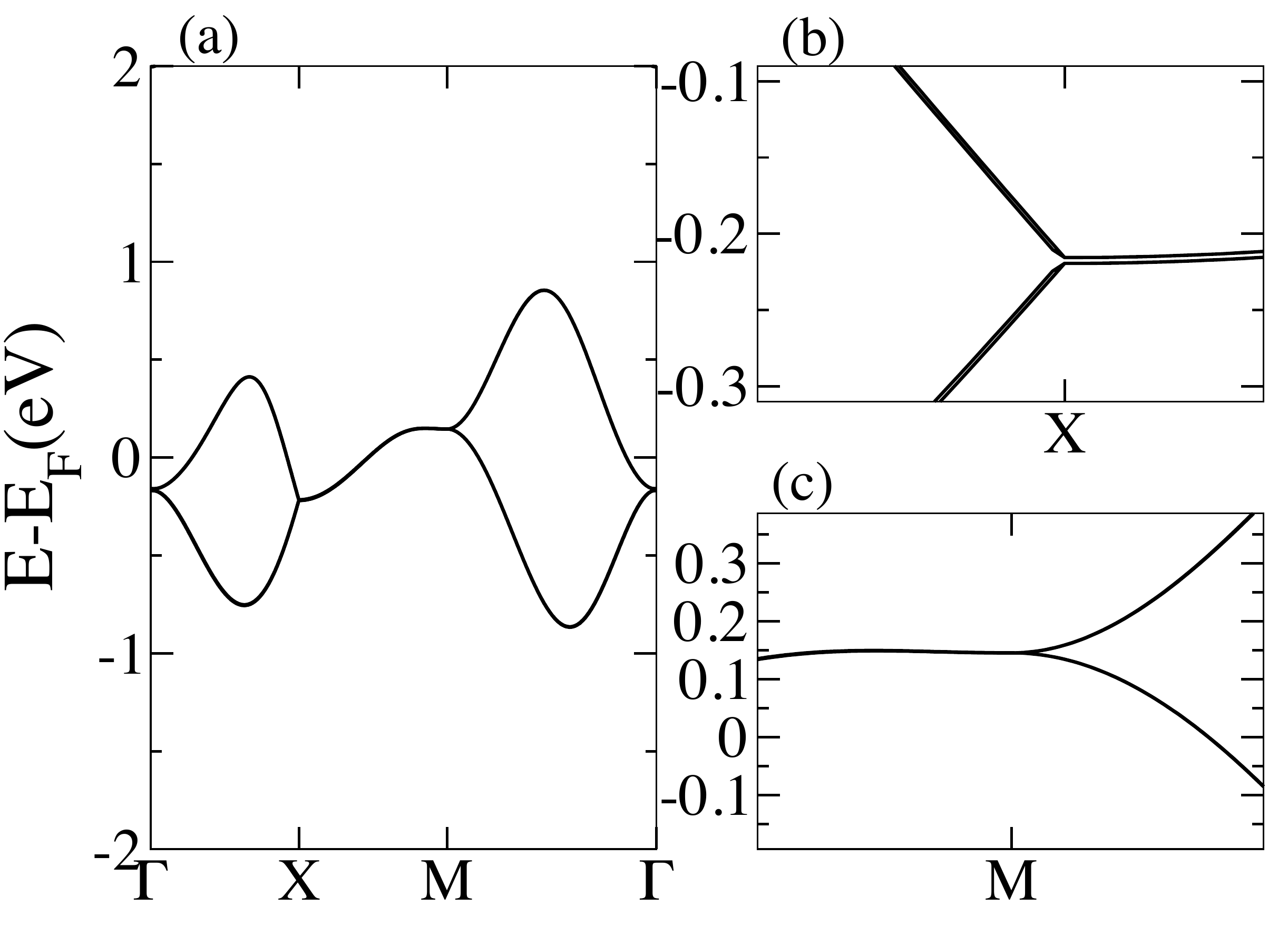}
 \caption{Penta-PC$_{2}$ band structure with SOC.}
  \label{SOC}
\end{figure}

 Since these theoretical materials preserve the PG space group $P\bar{4}2_{1}mm$, they show a similar trend in the electronic band structure. However, now there is a fractional
filling of the conduction (valence) band, which can be described in this
trivial phase (no SOC + TRS) by a single EBR. As stated by Bradlyn {\it et al.} \cite{natBradlyn2017}, if the Fermi level sits on a single EBR with fractional filling, the corresponding material is necessarily a protected (semi)metal. Thus, all these metal-PG and penta-XC$_{2}$  materials are symmetry-protected metals.
We have some remarks about this result. The above-mentioned symmetry protection is of crystalline character; 
since the space group is nonsymmorphic and TRS symmetry is preserved, it implies the well-known "stick-together" phenomenon for
energy bands \cite{DresselJorioBook} along the $Y$ line. This effect can be seen in the band structures of all penta-materials in this regime. It is remarkable that the sticky bands occur along all the BZ boundary. This implies that there is a trivial crystalline 
nodal line for these penta-materials. The nodal line presents a certain dispersion, i.e., it is not at constant energy in momentum space.
This effect is mainly induced by the lack of inversion symmetry which moves the nodal states to different energies, 
a fact that has been demonstrated in general in previous works \cite{Burkov2011,Young2015}.
For PG-adsorbed or substituted materials, although the nodal line is energy-dependent, it crosses the Fermi energy,
producing a single nodal point plus pockets of electrons or holes. This can be clearly seen in the band structures shown 
in Figs. \ref{pentamatbands1} and \ref{pentamatbands2}. Looking closer to the local low-energy behavior of the nodal line at the vicinity of the $Y$ line, we observe that bands have a linear dependence on $k_x$ along constant $k_y$ lines, so these carriers behave as massless fermions.  
This can be observed in a momentum space cut presented in Fig. \ref{nodal}. The massless fermion low-energy dispersion becomes more relevant if the Fermi level actually sits on a state of the nodal line. This crucially depends on the band filling fraction, being realizable in some of the materials studied.
Notice that, although we have a trivial phase in these materials, we still can have protected edge states. This can be explained in terms of the ten-fold way classification of the Fermi surface \cite{Matsuura2013}. As the considered materials belong to the AIII (chiral unitary) class, for spatial dimension $d=2$ a trivial phase arises, as expected. But due to an inherited non-triviality from another related AI (orthogonal) class, robust edge states that present linear or dispersionless characteristics might appear \cite{Chiu2016}.
 Next, SOC effects are explored by means of first-principles calculations. To this end, we use as an example penta-PC$_{2}$. Its band structure is presented in Fig. \ref{SOC}. As we are dealing with light elements the effect of SOC is rather weak; therefore, all these materials will behave as nodal line semimetals at room temperature, showing a continuum of Dirac nodes along the Brillouin zone boundary. Furthermore, these Dirac points are accessible under variations of the Fermi level position, making these massless fermions available under different perturbations, such as doping or electrostatic gating. 
 
Notwithstanding, the results derived by the symmetry analysis are confirmed. Namely, degeneracies at $\Gamma$ and $X$ points and along the $Y$ line are lifted. 
Likewise, the robustness of the $M$ point four-fold degeneracy is confirmed by these calculations, which allows us to identify this as a novel metallic phase similar to that  studied by Topp \textit{et al.} \cite{Topp2017}. 
The symmetry that protects the "stick-together" effect along the $Y$ line is broken, and the degeneracy of the above-mentioned high-symmetry points is also modified, implying the  disappearance of the nodal line. Therefore, for these penta-materials, SOC plus TRS enforces a transition from a nodal line metal state to a spin-orbit Dirac-node metal with nodal points located at $M$ \cite{Guan2017,Klemenz2018}, both phases being topologically trivial.

\subsection{Topological phase: Breaking TRS}

\begin{figure}
%[h!]
\centering
\includegraphics[width=1.0\columnwidth,clip]{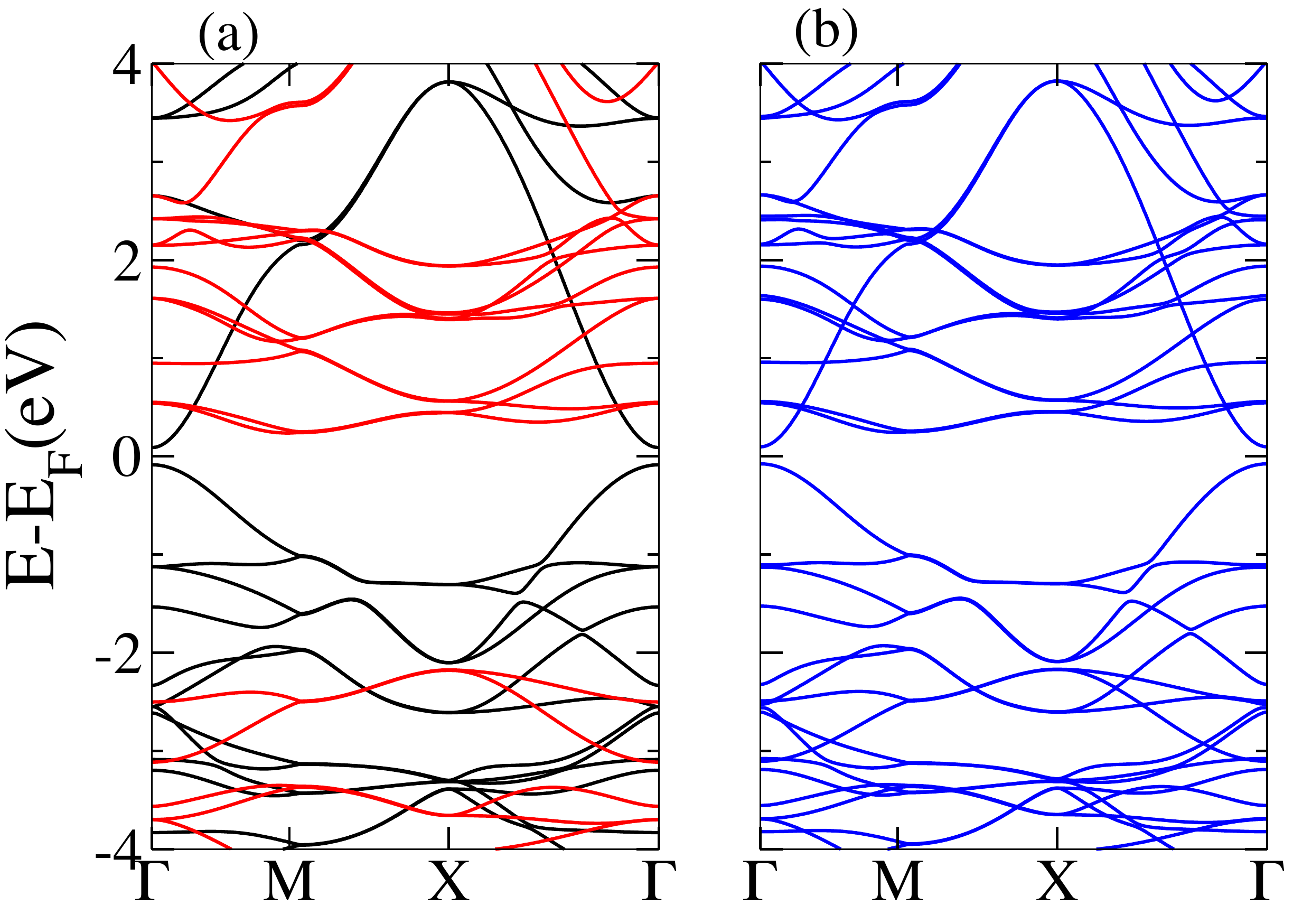}
 \caption{DFT penta-MnC2 band structure (a) without SOC  and (b) with SOC .}
  \label{Mn_PG}
\end{figure}

In what follows we show an example of a penta-material for the TRS-breaking phase with SOC: PG with substitution of Mn atoms at the 4e WP (or penta-MnC$_{2}$). In this case, magnetic Mn atoms break TRS and induce an intrinsic spin-orbit interaction, which results in a nontrivial phase. The corresponding band structure calculations are presented in Fig. \ref{Mn_PG}. 
In this example the four-fold to two-fold change of degeneracy at the $M$ point due to TRS breaking can be corroborated.  
We observe that along this high-symmetry path no band crossing among the four-band subgroups occurs. Further, the $Y$ line is completely nondegenerate, as expected. Thus, we have a situation similar to the $"a"$ phase presented in the model introduced in Sec. \ref{sec:topan}.
We also find  that the structure is magnetic and choose a ferromagnetic configuration, which is more energetically favorable. We ignore further magnetic group information for the subsequent analysis and restrict only to double space group data.    

In the case of non-trivial phases it is also necessary to establish a topological classification based on the calculation of topological invariants. In this case, we will use a numerical technique known as Wannier Charge Center (WCC) evolution  \cite{Taherinejad2014,Soluyanov2011}. In order to implement this procedure, an effective model defined in terms of Wannier functions must be supplied. We construct this model for the low-energy regime of penta-MnC$_{2}$ using the code Wannier90 \cite{Marzari2012,Mostofi2014}. This code uses a DFT band structure calculation as the input and wannierize the system by projecting the eigenfunction space to an initial set of orbitals. We chose  $sp^{3}$ orbitals for C atoms plus $s$ and $d$ orbitals for Mn atoms. This is an extension of the basic model with $p$ orbitals only, however, no d orbitals in this case transforms in a similiar form hat p orbital for the SSG of the 4e WP. 
We set a tolerance of $10^{-10}$ for the wannierization (minimization) procedure and define a frozen energy window of 3 eV around the Fermi level taking into account 20 bands.

 \begin{figure}[h!]
\centering
\includegraphics[width=1.0\columnwidth,clip]{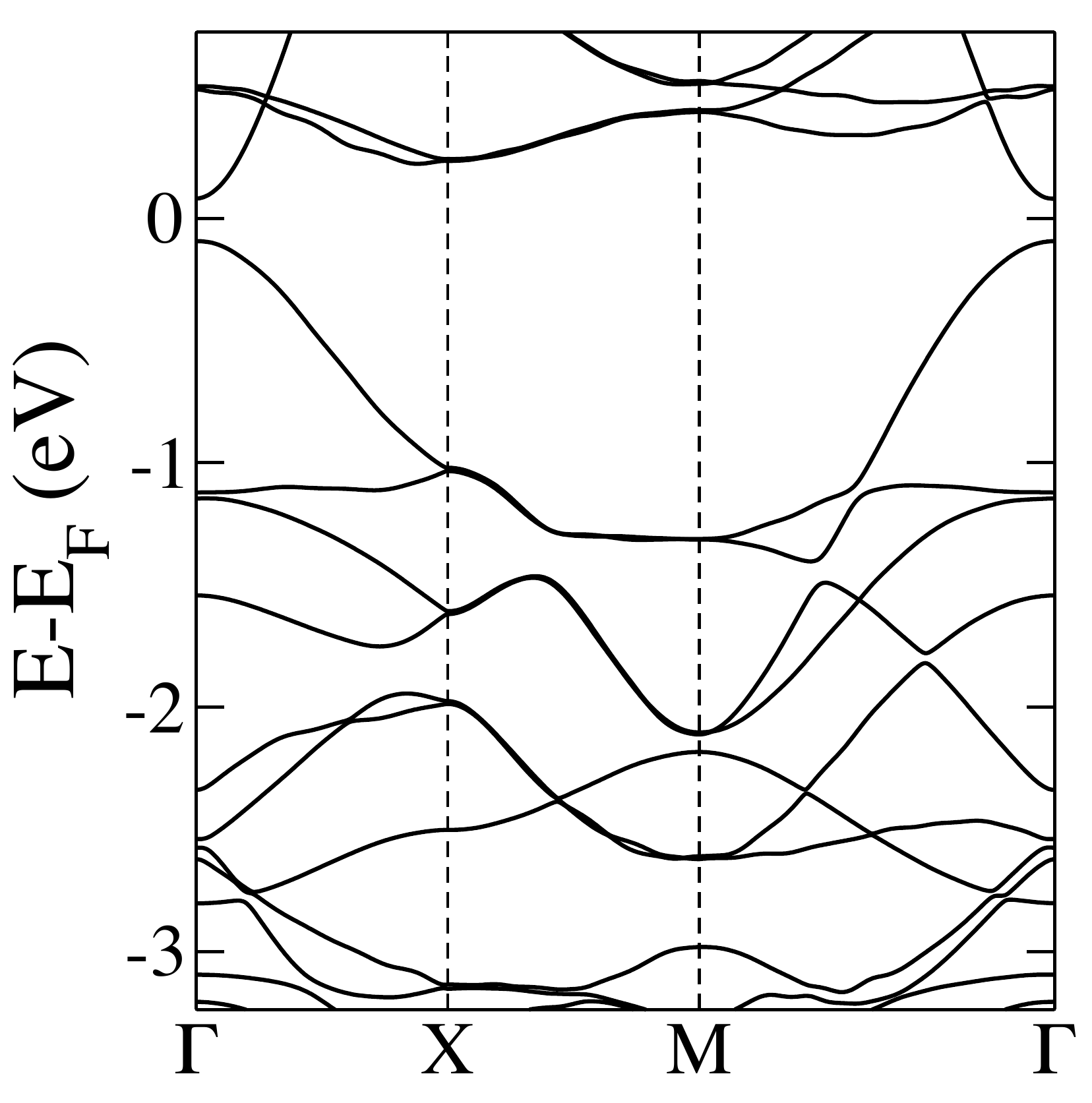}
 \caption{ Penta-MnC2 band structure calculated with the Wannier function basis.}
  \label{W90}
\end{figure}
With this model at hand, its band structure can be computed for penta-MnC$_{2}$; it is shown in Fig.\ref{W90}. It agrees very well with that obtained from first-principles methods, being a good starting model for the topological invariant calculations. 

\begin{figure}[h!]
\centering
\includegraphics[width=1.0\columnwidth,clip]{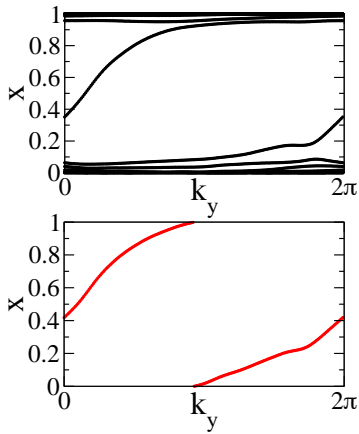}
\caption{Top: WCC evolution for $x$ axis. Bottom: Sum of WCC for the $x$ axis.}
 \label{Wcc_x}
\end{figure}

\begin{figure}[h!]
\centering
\includegraphics[width=1.0\columnwidth,clip]{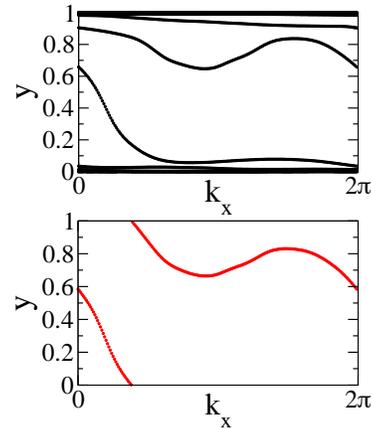}
 \caption{Top: WCC evolution for $y$ axis. Bottom: Sum of WCC for the $y$ axis.}
  \label{Wcc_y}
\end{figure}

%As mentioned before, the numerical computation of topological invariants can be achieved by Wannier charge center evolution  \cite{Taherinejad2014,Soluyanov2011}.
The WCC evolution is related to several topological invariants such as the Z2 invariant and Chern numbers. In the particular case we are studying, since TRS is broken \cite{Bernevig2006} we  just compute Chern numbers. This is done by taking the sum of the WCCs at each $k$ point and then counting the times this function winds across the vertical axis while along the entire $k$-path \cite{Taherinejad2014}. This procedure has the advantage that only bulk properties are needed to compute the topological invariants of a material, without resorting to more expensive surface or edge state calculations.   
To this purpose, we have used two codes: Z2pack \cite{Gresch2017} and WannierTools \cite{Wu2018}. Only the Z2pack results are presented here, since WannierTools gives a similar outcome. Z2pack uses the Hamiltonian in terms of Wannier functions as an input. Additionally, we set a position tolerance of 0.001, a gap tolerance equal to 0.1 eV and a number of lines of 200. The WCC calculation can be carried for each one of the spatial directions of the system. For penta-materials, being 2D systems, there are two directions, $x$ and $y$. We present the WCC evolution and WCC sum for both directions in Figs. \ref{Wcc_x} and \ref{Wcc_y}, respectively.

It can be observed that  the respective Chern numbers are non-zero for both directions. In particular, a Chern number of $C_{x}=1$ was obtained for the $x$-direction and $C_{y}=-1$ for the $y$-direction, related to the winding of the sum of the WCC across the respective $k$ directions. This implies a charge pumping effect along each $k$-space direction \cite{Soluyanov2011b,Gresch2017} and hints for a nontrivial topological phase at the bulk gap.
If we additionally define a global Chern number as the sum of $C_{x}$ and $C_{y}$, we obtain zero. This indicates that a direction-dependent protection of robust states is present in the system, which can be related to a crystalline symmetry protection \cite{Taherinejad2014}. This symmetry can be tracked to an axis-dependent operation, which in this case corresponds to a nonsymmorphic $\pi$-rotation about either the $ x$  or $y$ axis plus a fractional translation \cite{MillerLoveBook}. Thus we conclude that penta-MnC$_{2}$ is a material realization for a magnetic topological insulator phase.

In summary, we have studied a set of materials sharing a space group structure that we named penta-materials. A detailed analysis for many physical possibilities, including TRS and TRS breaking as well as SOC, shows that these materials can host different phases. We have found nodal line fermions if SOC is weak and TRS is present; in metallic penta-materials; this implies that a continuum of Dirac points are accessible around the Fermi energy. Also, symmetry-protected Dirac points arise if SOC is strong enough. Finally, if TRS is broken and SOC is sufficiently high, we encounter a nontrivial topological phase characterized by non-zero Chern numbers for both primitive directions in  $k$-space. 
%Further work is needed in order to fully characterize 
%these    
%phases in 
A wider class of penta-materials, from other substitutions or functionalization, awaits for the full characterization of their topological phases and 
%to obtain 
the obtention of 
additional physical properties.    

\acknowledgments
This work has been partially supported by Chilean CONICYT PhD scholarship No. 21150492,  Chilean FONDECYT Grant No.1151316 and the Spanish MINECO and the European Union under Grant No. FIS2015-64654 P/MINECO/FEDER, CSIC i-coop Grant No. ICOOPA-20150. J.D.C. thanks the Laboratorio de Simulaci\'on y Computaci\'on Cient\'{i}fica at Universidad de Medell\'{i}n for computational time.

\appendix
\section{Lattice constants}
\label{latt}
\counterwithin{table}{section}
\setcounter{table}{0}

Here we present a summary chart for the first-principles geometrical parameters of penta-materials studied in this work. Results were obtained from SIESTA code.

\begin{table}[h!]
\caption{{\label{tableA1}} Here $a$ is the lattice constant, $cc$ the carbon-carbon distance, $ cx$   the distance between  a carbon atom and other element $x$ and  $d$ is the distance between the carbon atom and the absorbed Li or Na. }
%
%\begin{center}
%\renewcommand{\arraystretch}{1.5}
\begin{tabular}{|c|c|c|c|c|}
\hline 
System & $a(\text{\AA)}$ & $cc(\text{\AA)}$ & $cx(\text{\AA)}$ & $d(\text{\AA)}$\tabularnewline
\hline 
\hline 
PG & 3.650 & 1.554 & & \tabularnewline
\hline 
NC$_{2}$ & 3.417 & 1.403 & 1.516 & \tabularnewline
\hline 
PC$_{2}$ & 4.178 & 1.418 & 1.815 & \tabularnewline
\hline 
BC$_{2}$ & 4.017 & 1.388 & 1.647 & \tabularnewline
\hline 
MnC$_{2}$ & 4.630 & 1.262 & 2.183 & \tabularnewline
\hline 
Li-PG & 3.695 & 1.576 & & 2.188\tabularnewline
\hline 
Na-PG & 3.790 & 1.602 & & 2.564\tabularnewline
\hline 
\end{tabular}
\end{table}

\section{Character tables}
\label{chartab}
\counterwithin{table}{section}
\setcounter{table}{0}

Character tables for the double groups of the wavevector at different points and lines of the Brillouin zone are presented. The character information and the notation for the symmetry operations is adapted from the Bilbao crystallographic server \cite{Bilbao}.

\begin{table}[h!]
\caption{{\label{tableB1}}Character table for double space group no. 113 and $\Gamma$ point.}
%
%\begin{center}
\renewcommand{\arraystretch}{1.3}
\begin{tabular}{@{\hspace{1em}}c@{\hspace{1em}}c@{\hspace{1em}}cccc@{\hspace{1em}}c@{\hspace{1em}}c@{\hspace{1em}}}
\toprule
 & $I$ & $2_{001}$ & $\overline{4}^{+}_{001}$ & $(2_{010}|\frac{1}{2}\frac{1}{2}0)$ & $(m_{110}|\frac{1}{2}\frac{1}{2}0)$ & $I^{d}$ & $\overline{4}^{+d}_{001}$\\
 \colrule
$\Gamma_{1}$ & 1 & 1 & 1 & 1 & 1 & 1 & 1\\
$\Gamma_{2}$ & 1 & 1 & $-1$ & 1 & $-1$ & 1 & $-1$\\
$\Gamma_{3}$ & 1 & 1 & $-1$ & $-1$ & 1 & 1 & $-1$\\
$\Gamma_{4}$ & 1 & 1 & 1 & $-1$ & $-1$ & 1 & 1\\
$\Gamma_{5}$ & 2 & $-2$ & 0 & 0 & 0 & 2 & 0\\
$\bar{\Gamma}_{6}$ & 2 & 0 & $-\sqrt{2}$ & 0 & 0 & $-2$ & $\sqrt{2}$\\
$\bar{\Gamma_{7}}$ & 2 & 0 & $\sqrt{2}$ & 0 & 0 & $-2$ & $-\sqrt{2}$\\
\botrule
\end{tabular}
%\end{center}
\end{table}

%\FloatBarrier
\begin{table*}
\caption{Character table for the double group at X point. }
%\begin{minipage}{0.7\textwidth}
%\renewcommand{\arraystretch}{1.3}
\begin{ruledtabular}
\begin{tabular}{ccccccccc}
 & $I$ & $2_{001}$ & $(2_{010}|\frac{1}{2}\frac{1}{2}0)$ & $(2_{100}|\frac{1}{2}\frac{1}{2}0)$ & $I^{d}$ & $2^{d}_{001}$ & $(2^{d}_{010}|\frac{1}{2}\frac{1}{2}0)$ & $(2^{d}_{100}|\frac{1}{2}\frac{1}{2}0)$\\
 \hline
$X_{1}$ & 2 & 0 & 0 & 0 & 2 & 0 & 0 & 0\\
$\bar{X}_{2}$ & 1 & $-i$ & 1 & $-i$ & $-1$ & $i$ & $-1$ & $i$\\
$\bar{X}_{3}$ & 1 & $i$ & $-1$ & $-i$ & $-1$ & $-i$ & 1 & $i$\\
$\bar{X}_{4}$ & 1 & $-i$ & $-1$ & $i$ & $-1$ & $i$ & 1 & $-i$\\
$\bar{X}_{5}$ & 1 & $i$ & 1 & $i$ & $-1$ & $-i$ & $-1$ & $-i$\\
\end{tabular}
\end{ruledtabular}
\label{tableB2}
%\end{center}
%\end{minipage}
\end{table*}

\begin{table*}
\begin{ruledtabular}
\begin{tabular}{ccccc}
 & $I$ & $2_{001}$ & $I^{d}$ & $(2^{d}_{100}|\frac{1}{2}\frac{1}{2}0)$\\
\hline 
$Y_{1}$ & 1 & $e^{i\pi u}$ & 1 & $e^{i\pi u}$\\
$Y_{2}$ & 1 & $e^{i\pi(1+u)}$ & 1 & $e^{i\pi(1+u)}$\\
$\bar{Y}_{3}$ & 1 & $e^{-i\pi(\frac{1}{2}-u)}$ & $-1$ & $e^{i\pi(\frac{1}{2}+u)}$\\
$\bar{Y}_{4}$ & 1 & $e^{i\pi(\frac{1}{2}+u)}$ & $-1$ & $e^{-i\pi(\frac{1}{2}-u)}$\\
\end{tabular}
\end{ruledtabular}
\caption{Character table for the double group at Y line. Where $\boldsymbol{k}_{Y}=(\frac{1}{2},u,0)$ with $u\in(0,\frac{1}{2})$.}
\label{tableB3}
\end{table*}

%\FloatBarrier
\begin{table*}
%\begin{minipage}{0.8\textwidth}
%\renewcommand{\arraystretch}{1.3}
\begin{ruledtabular}
\begin{tabular}{ccccccccc}
 & $I$ & $2_{001}$ & $\overline{4}^{+}_{001}$ & $\overline{4}^{-}_{001}$ & $(2_{010}|\frac{1}{2}\frac{1}{2}0)$ & $(2_{100}|\frac{1}{2}\frac{1}{2}0)$ & $(m_{110}|\frac{1}{2}\frac{1}{2}0)$ & $(m_{1\overline{1}0}|\frac{1}{2}\frac{1}{2}0)$\\ 
 \hline 
$M_{1}$ & 1 & $-1$ & $i$ & $-i$ & $i$ & $-i$ & 1 & $-1$\\
$M_{2}$ & 1 & $-1$ & $-i$ & $i$ & $i$ & $-i$ & $-1$ & 1\\
$M_{3}$ & 1 & $-1$ & $-i$ & $i$ & $-i$ & $i$ & 1 & $-1$\\ 
$M_{4}$ & 1 & $-1$ & $i$ & $-i$ & $-i$ & $i$ & $-1$ & 1\\
$M_{5}$ & 2 & 2 & 0 & 0 & 0 & 0 & 0 & 0\\
$\bar{M}_{6}$ & 2 & 0 & $\sqrt{2}i$ & $-\sqrt{2}i$& 0 & 0 & 0 & 0\\ 
$\bar{M}_{7}$ & 2 & 0 & $-\sqrt{2}i$ & $\sqrt{2}i$ & 0 & 0 & 0 & 0\\ 
\end{tabular}
\end{ruledtabular}
%\end{minipage}
\caption{\label{tableB4}Character table for the double group at M point. }
\end{table*}
%caption

\begin{table*}[h!]
%[b]
%\begin{minipage}{0.8\textwidth}
%\renewcommand{\arraystretch}{1.3}
\begin{ruledtabular}
\begin{tabular}{ccccccccc}
 & $I^{d}$ & $2^{d}_{001}$ & $\overline{4}^{+d}_{001}$ & $\overline{4}^{-d}_{001}$ & $(2^{d}_{010}|\frac{1}{2}\frac{1}{2}0)$ & $(2^{d}_{100}|\frac{1}{2}\frac{1}{2}0)$ & $(m^{d}_{110}|\frac{1}{2}\frac{1}{2}0)$ & $(m^{d}_{1\overline{1}0}|\frac{1}{2}\frac{1}{2}0)$\\
\hline 
$M_{1}$ & 1 & $-1$ & $i$ & $-i$ & $-i$ & $i$ & 1 & $-1$\\
$M_{2}$ & 1 & $-1$ & $-i$ & $i$ & $-i$ & $i$ & $-1$ & 1\\
$M_{3}$ & 1 & $-1$ & $-i$ & $i$ & $i$ & $-i$ & 1 & $-1$\\
$M_{4}$ & 1 & $-1$ & $i$ & $-i$ & $i$ & $-i$ & $-1$ & 1\\
$M_{5}$ & 2 & 2 & 0 & 0 & 0 & 0 & 0 & 0\\
$\bar{M}_{6}$ & $-2$ & 0 & $-\sqrt{2}i$ & $\sqrt{2}i$ & 0 & 0 & 0 & 0\\
$\bar{M}_{7}$ & $-2$ & 0 & $\sqrt{2}i$ & $-\sqrt{2}i$ & 0 & 0 & 0 & 0\\
\end{tabular}
\end{ruledtabular}
%\end{minipage}
\caption{\label{tableB5} Character table for the double group at M point (continuation).}
\end{table*}

\FloatBarrier
%\bibliography{ref_penta3.bib}

%

\end{document}